\documentclass[conference]{IEEEtran}
\IEEEoverridecommandlockouts

\ifdefined\pdfminorversion
\else
  \ifdefined\pdfvariable
    \pdfvariable minorversion=7
  \fi
\fi

\usepackage[pdftex]{graphicx}
\graphicspath{{graphics/}{./}}
\DeclareGraphicsExtensions{.pdf,.jpeg,.png}

\usepackage[utf8]{inputenc} \usepackage[T1]{fontenc}    \usepackage{url}            \usepackage{microtype}      \usepackage[dvipsnames]{xcolor}

\usepackage{amsmath,amssymb,amsfonts}
\usepackage{mathtools}
\usepackage{bm}

\usepackage{array}
\usepackage{booktabs}       \usepackage{multirow}
\usepackage{threeparttable}

\usepackage{subcaption}
\usepackage{algorithm}
\usepackage{algpseudocode}  

\usepackage{csquotes}
\usepackage{xpatch}
\usepackage[backend=biber,                          style=ieee,                             ]{biblatex}
\DeclareFieldFormat{doi}{\mkbibacro{DOI}\addcolon\space
  \ifhyperref
    {\href{https://doi.org/#1}{\nolinkurl{#1}}}
    {\nolinkurl{#1}}}
\usepackage[automake, nomain, acronym, xindy, shortcuts=ac]{glossaries-extra}
\glsdisablehyper
\makeglossaries
\setabbreviationstyle[acronym]{long-short} \glssetcategoryattribute{acronym}{nohyperfirst}{true}
\newacronym{tb}{TB}{tuberculosis}
\newacronym{ptb}{PTB}{pulmonary tuberculosis}
\newacronym{lus}{LUS}{lung ultrasound}
\newacronym{pocus}{POCUS}{point-of-care ultrasound}
\newacronym{fash}{FASH}{focused assessment with sonography for HIV-associated tuberculosis}
\newacronym{who}{WHO}{World Health Organization}
\newacronym{ultrai}{ULTR-AI}{ultrasound-led tuberculosis recognition using artificial intelligence}
\newacronym{cad}{CAD}{computer-aided detection}
\newacronym{cadlus4tb}{CAD LUS4TB}{computer-aided detection using lung ultrasound for tuberculosis}
\newacronym{cxr}{CXR}{chest radiograph}
\newacronym{tpp}{TPP}{target product profile}
\newacronym{ssl}{SSL}{self-supervised learning}
\newacronym{mae}{MAE}{masked autoencoder}
\newacronym{jepa}{JEPA}{joint-embedding predictive architecture}
\newacronym{vit}{ViT}{vision transformer}
\newacronym{vitl}{ViT-L}{large vision transformer}
\newacronym{cls}{CLS}{classification token}
\newacronym{cnn}{CNN}{convolutional neural network}
\newacronym{resnet}{ResNet}{residual neural network}
\newacronym{mil}{MIL}{multiple-instance learning}
\newacronym{mlp}{MLP}{multi-layer perceptron}
\newacronym{rope}{RoPE}{rotary position embedding}
\newacronym[longplural={areas under the receiver operating characteristic curve}]{auc}{AUROC}{area under the receiver operating characteristic curve}
\newacronym{auprc}{AUPRC}{area under the precision-recall curve}
\newacronym{roc}{ROC}{receiver operating characteristic}
\newacronym{rvsp}{RVSP}{right ventricular systolic pressure}
\newacronym{ci}{CI}{confidence interval}
\newacronym{dicom}{DICOM}{digital imaging and communications in medicine}
\newacronym{roi}{ROI}{region of interest}
\newacronym{hiv}{HIV}{human immunodeficiency virus}
 
\usepackage{siunitx}
\DeclareSIUnit{\unitless}{\relax}
\DeclareSIUnit{\px}{px}

\usepackage{hyperref}       \usepackage{textcomp}
\usepackage{etoolbox}
\usepackage[capitalise, noabbrev, nameinlink]{cleveref}

\usepackage[normalem]{ulem}
\robustify\bfseries
\robustify\uline
\robustify\uuline
\robustify\tnote

\def\BibTeX{{\rm B\kern-.05em{\sc i\kern-.025em b}\kern-.08em
    T\kern-.1667em\lower.7ex\hbox{E}\kern-.125emX}}

\newcommand{\emailaddr}[1]{{\footnotesize\normalfont\texttt{#1}}}

\newcommand{\subtitle}[1]{\\[-0.12\baselineskip]{\normalfont\large #1}}

\newtoggle{blindreview}
\togglefalse{blindreview}
\ifdef{\BLINDREVIEW}{\toggletrue{blindreview}}{}
\ifdef{\CAMERAREADY}{\togglefalse{blindreview}}{}
\iftoggle{blindreview}
  {\typeout{BLIND mode: review copy}}
  {\typeout{BLIND mode: camera-ready}}

\usepackage[switch]{lineno}
\newtoggle{linenumbers}
\togglefalse{linenumbers}

\iftoggle{blindreview}{\newcommand{\annotation}[3]{}}{\newcommand{\annotation}[3]{\textcolor{#1}{\scriptsize \slshape {[#2: #3]}}}}

\newsavebox{\hiddenbox}
\newlength{\hiddenwd}
\newlength{\hiddenht}
\newlength{\hiddendp}
\newlength{\hiddenindent}
\newenvironment{hidden}[1][\textit{Withheld for purpose of blind peer-review}]{\def\hiddennotice{#1}\setlength{\hiddenindent}{\parindent}\begin{lrbox}{\hiddenbox}\begin{minipage}[t]{\linewidth}\setlength{\parindent}{\hiddenindent}}{\end{minipage}\end{lrbox}\setlength{\hiddenwd}{\wd\hiddenbox}\setlength{\hiddenht}{\ht\hiddenbox}\setlength{\hiddendp}{\dp\hiddenbox}\iftoggle{blindreview}{\sbox{\hiddenbox}{\parbox[t]{\hiddenwd}{\centering\hiddennotice}}\ifdim\ht\hiddenbox<\hiddenht \ht\hiddenbox\hiddenht \fi
    \ifdim\dp\hiddenbox<\hiddendp \dp\hiddenbox\hiddendp \fi
    \ifdim\dimexpr\ht\hiddenbox+\dp\hiddenbox\relax
         >\dimexpr\hiddenht+\hiddendp\relax
      \PackageWarning{blindreview}{Notice does not fit the space it replaces}\fi
  }{}\par\noindent\usebox{\hiddenbox}\par
}

\makeatletter
\newsavebox{\blindtitlea}
\newsavebox{\blindtitleb}
\newlength{\blinddelta}

\newcommand{\blindmeasure}[2]{\global\setbox#1\vbox{\hsize\textwidth \@parboxrestore \col@number\@ne
    \normalfont\normalsize
    \hbadness\@M \vbadness\@M \hfuzz\maxdimen \vfuzz\maxdimen
    \IEEEquantizedisabletitlecmds
    \let\@author#2\@maketitle}}

\newcommand{\reserveauthorspace}{\blindmeasure\blindtitlea\authorreal
  \blindmeasure\blindtitleb\authorblind
  \setlength{\blinddelta}{\dimexpr\ht\blindtitlea-\ht\blindtitleb\relax}\ifdim\blinddelta<\z@
    \PackageWarning{blindreview}{Notice is taller than the author block}\else
    \setlength{\blindpadabove}{0.5\blinddelta}\setlength{\blindpadbelow}{\dimexpr\blinddelta-\blindpadabove\relax}\blindmeasure\blindtitleb\authorblind
    \addtolength{\blindpadbelow}{\dimexpr\ht\blindtitlea-\ht\blindtitleb\relax}\fi
  \blindmeasure\blindtitleb\authorblind
  \typeout{BLIND title height: real=\the\ht\blindtitlea\space
    blind=\the\ht\blindtitleb\space reserved=\the\blinddelta}\ifdim\ht\blindtitlea=\ht\blindtitleb\else
    \PackageWarning{blindreview}{Title height mismatch}\fi
}
\makeatother

\newcommand{\pp}[1]{\num[round-precision=2, round-pad=true,
                         drop-zero-decimal=false]{#1}}

\newcolumntype{I}[1]{@{\,[}S[#1]@{,\;}S[#1]@{]\hspace{2\tabcolsep}}}
\newcolumntype{J}[1]{@{\,[}S[#1]@{,\;}S[#1]@{]}}

\begin{document}

\iftoggle{linenumbers}{\linenumbers}{}

\title{Wrong Organ, Right Physics\subtitle{Transferring Echocardiography Pretraining to Lung Ultrasound for
Tuberculosis Screening}}

\newcommand{\orcidauthor}[2]{#1\hspace*{0.6mm}\href{https://orcid.org/#2}{\includegraphics[width=0.9em]{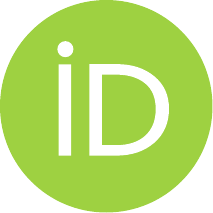}}}

\newcommand{\affilsun}{\textit{Department of Electrical and Electronic Engineering},
  \textit{University of Stellenbosch}, Stellenbosch, South Africa}
\newcommand{\affilepfl}{\textit{Laboratory for intelligent Global Health and Humanitarian Response
  Technologies (LiGHT)}\\
  \textit{Swiss Federal Institute of Technology (EPFL)}, Lausanne, Switzerland}
\newcommand{\affilctr}{\textit{South African Medical Research Council Centre for Tuberculosis
  Research (CTR)}\\
  \textit{Division of Molecular Biology and Human Genetics, Faculty of Medicine
  and Health Sciences}\\
  \textit{Stellenbosch University}, Cape Town, South Africa}
\newcommand{\affilcnhu}{\textit{National Teaching Hospital for Tuberculosis and Pulmonary Diseases
  (CNHU-PPC)}, Cotonou, Benin}

\newcommand{\authorreal}{\IEEEauthorblockN{\orcidauthor{Christiaan M. Geldenhuys}{0000-0003-0691-0235}\IEEEauthorrefmark{1},
  \orcidauthor{Joshua M. Jansen van Vüren}{0000-0002-3406-4788}\IEEEauthorrefmark{1},
  \orcidauthor{Véronique Suttels}{0000-0002-5261-0439}\IEEEauthorrefmark{2},
  \orcidauthor{Trevor Brokowski}{0000-0002-7547-7749}\IEEEauthorrefmark{2},\\
  \orcidauthor{Ablo P. Wachinou}{0000-0002-1024-9665}\IEEEauthorrefmark{4},
  \orcidauthor{Mary-Anne Hartley}{0000-0002-8826-3870}\IEEEauthorrefmark{2},
  \orcidauthor{Rensu P. Theart}{0000-0003-0508-1690}\IEEEauthorrefmark{1},
  \orcidauthor{Grant Theron}{0000-0002-9216-2415}\IEEEauthorrefmark{3},
  \orcidauthor{Thomas R. Niesler}{0000-0002-7341-1017}\IEEEauthorrefmark{1}}\IEEEauthorblockA{\IEEEauthorrefmark{1}\affilsun}\IEEEauthorblockA{\IEEEauthorrefmark{2}\affilepfl}\IEEEauthorblockA{\IEEEauthorrefmark{4}\affilcnhu}\IEEEauthorblockA{\IEEEauthorrefmark{3}\affilctr}\IEEEauthorblockA{{\footnotesize\normalfont Corresponding authors:}\hspace*{0.35em}\emailaddr{cmgeldenhuys@sun.ac.za}, \emailaddr{trn@sun.ac.za}}}

\newlength{\blindpadabove}
\newlength{\blindpadbelow}
\newcommand{\authorblind}{\IEEEauthorblockN{\\[\blindpadabove]\textbf{\texttt{Authors omitted for purpose of blind peer-review}}\\[\blindpadbelow]}}

\author{\authorreal}
\iftoggle{blindreview}{\reserveauthorspace\author{\authorblind}}{}

\hypersetup{
pdftitle={Wrong Organ, Right Physics: Transferring Echocardiography Pretraining to Lung Ultrasound for Tuberculosis Screening},
pdfkeywords={lung ultrasound, tuberculosis screening, echocardiography, transfer learning, pretrained encoders, feature standardisation, model selection, small-cohort evaluation, joint-embedding predictive architecture},
}
\iftoggle{blindreview}{}{\hypersetup{pdfauthor={Christiaan M.~Geldenhuys,Joshua M.~Jansen van Vüren,Véronique Suttels,Trevor Brokowski,Ablo P.~Wachinou,Mary-Anne Hartley,Rensu P. Theart,Grant Theron,Thomas R.~Niesler}}}
\maketitle

\begin{abstract}
	\Ac{lus} is attractive for \ac{tb} screening at primary-care level, but labelled cohorts are small.
	Echocardiography carries no such constraint, while sharing the same underlying ultrasound imaging physics, signal processing and B-mode appearance as \ac{lus}.
	We ask whether an encoder pretrained on that high-resource ultrasound domain carries representations that remain usable in the low-resource one.
	Only the encoder varies, across seventeen encoders spanning three architecture families.
	Among them, a latent-predictive video encoder pretrained on generic video (V-JEPA2-L) and its echocardiography counterpart (EchoJEPA-L) differ in pretraining corpus alone.
	The choice among these encoders does not resolve the classification, the whole family spanning \pp{2.50} percentage points against a measurement resolution of \pp{2.71}.
	What moves the task instead is feature conditioning.
	Standardising the features between the encoder and the classifier improves all seventeen encoders by a mean of \pp{+1.23} percentage points at $p = \num{1.5e-5}$.
	On the held-out test set every encoder selected on the development folds stands above the baseline system by up to \pp{+2.57} percentage points of \ac{auc}, and specificity at \qty{90}{\percent} sensitivity reaches \qty{79.3}{\percent} against \qty{60.3}{\percent}.
	The contrast specified in advance, EchoJEPA-L against V-JEPA2-L, measures \pp{-0.16} percentage points at $p = \num{0.926}$.
	We therefore find no evidence that shared ultrasonic physics alone makes echocardiography a more productive pretraining corpus than generic video, and any advantage, if present, is smaller than this cohort can resolve.
	The video encoders receive replicated still images, however, so whether this absence of an effect reflects the pretraining domain or a video encoder applied to static frames cannot be separated.
	The limiting factor is the labelled cohort rather than the encoder.
\end{abstract}

\vspace*{1mm}

\glsresetall

\section{Introduction}
\label{sec:intro}

\Ac{tb} remains one of the leading infectious causes of death.
An estimated one in four people who fell ill with TB in 2023 was never diagnosed~\autocite{qin2025cadvsradiologists}.
Chest radiography, interpreted either by a trained reader or by \ac{cad} software, is the imaging tool the \ac{who} recommends for screening and triage.
Yet radiographic equipment and trained readers remain concentrated in urban tertiary centres \autocite{who2021tbscreeninghandbook}.
\Ac{lus} is an attractive complement at primary-care level, because the probe is portable, battery-powered and free of ionising radiation.
The weakness of \ac{lus} lies in interpretation, which is operator-dependent and not yet supported by a strong body of evidence~\autocite{bigio2021pocustbreview}.

Automated interpretation can address this core weakness, but labelled \ac{lus} data are scarce.
The cohort studied here includes 504 labelled patients, of which 101 are reserved as a held-out test partition.
This is small by the standards of modern representation learning.
Echocardiography, on the other hand, has public corpora that are several orders of magnitude larger.
Furthermore, it is acquired under the same ultrasound imaging physics, and shares the speckle statistics and the B-mode display of \ac{lus}.

Transfer into medical imaging has usually drawn on natural images \autocite{raghu2019transfusion}, or on the target anatomy itself \autocite{vanberlo2023lusssl}.
Echocardiography offers a third source, one that shares the imaging physics of \ac{lus} but not its anatomy.
EchoJEPA is a latent-predictive encoder pretrained on echocardiography \autocite{munim2026echojepa}.
In this paper we exploit the asymmetry between the two domains.
We ask whether an encoder pretrained on a high-resource ultrasound domain carries representations that remain useful in the low-resource \ac{lus} domain, or whether the different anatomy, probe placement and acquisition protocol prevent this.

As a baseline, we use a classifier built on an 18-layer \gls{resnet} trained in a supervised way, which was found to perform well in previous work~\autocite{ultrai2025escmid}.
As competing encoders, we evaluate EchoJEPA-L against V-JEPA2-L, a video encoder of identical architecture pretrained on generic video and never on ultrasound.
The two therefore differ in pretraining corpus alone.
We further consider fifteen other encoders, spanning convolutional and transformer architectures and four pretraining corpora.
In our evaluation only the encoder varies, while the data, the classifier architecture and the evaluation protocol remain the same.
Each encoder is evaluated with its parameters held fixed, so that only the attention-pooling classifier is trained.

\subsection{Contributions}
\label{sec:contributions}
\begin{itemize}
	\item A test of whether shared imaging physics, without shared anatomy, makes echocardiography a productive pretraining corpus for \ac{lus}, using two encoders that differ in pretraining corpus alone (see \cref{sec:discussion:rq1}).

	\item A comparison of encoder architectures and pretraining corpora for tuberculosis classification from \ac{lus}, obtained under a protocol in which the encoder alone is exchanged (\cref{sec:results:encoders}).

	\item The isolation of feature conditioning as a variable separable from the choice of encoder, tested by pairing each encoder against itself (\cref{sec:results:norm}).
\end{itemize}

\section{Background}
\label{sec:background}

\subsection{Lung ultrasound for tuberculosis screening}
\label{sec:background:tb}

The interpretation of \ac{lus} images is a central limitation of the technique.
A systematic review of \acl{pocus} for \ac{ptb} found no consensus on how images should be acquired or analysed, and concluded that its diagnostic utility remains unclear \autocite{bigio2021pocustbreview}.
Effective automatic analysis would address this, filling a role similar to automated chest-radiograph analysis without the infrastructure or radiation control that even ultra-portable radiographic units require \autocite{alege2025ultraportablecad4tb}.

Applied as a screening test, \ac{lus} would take the place of a symptom questionnaire, so that every patient who presents is scanned rather than questioned.
Such a test falls under Type~2 of the 2025 \ac{who} \ac{tpp} for \ac{tb} screening tests, which supersedes the 2014 triage profile \autocite{who2014tpp,who2025screeningtpp}.
A Type~2 test screens at high sensitivity and is followed by a second screening test before a patient is referred for diagnostic evaluation.

\subsection{Latent-predictive self-supervised video representation learning}
\label{sec:background:jepa}

Masked \acl{ssl} divides broadly according to where in the network the loss is computed.
Reconstruction methods place it in pixel space, masking patches and regressing the missing values, as the \gls{mae} does for images \autocite{he2021mae} and VideoMAE for video \autocite{tong2022videomae}.
Such methods assume that local pixel intensities reflect the underlying structure.
For ultrasound that assumption is weak.
The image is dominated by stochastic speckle, and by acoustic shadow and attenuation that vary with probe angle and gain without corresponding to the anatomy.
A pixel-reconstruction objective must reproduce these artefacts faithfully to lower its loss, spending capacity on detail that does not carry across operators or devices \autocite{munim2026echojepa,radhachandran2026usjepa}.

Latent prediction moves the loss into the representation space instead.
\Textcite{lecun2022pathami} argues that an encoder trained under a \gls{jepa} objective can discard input detail that is not predictable, whereas a generative architecture cannot.
I-JEPA realises this for images \autocite{assran2023ijepa}, V-JEPA establishes feature prediction as a stand-alone video objective \autocite{bardes2024vjepa}, and V-JEPA~2 scales it to more than one million hours of video \autocite{assran2025vjepa2}.
EchoJEPA applies the pretraining procedure of V-JEPA~2 to echocardiography rather than to generic video \autocite{munim2026echojepa}.

\subsection{Large-scale pretrained ultrasound and echocardiography encoders}
\label{sec:background:usfm}

Echocardiography already has several large-scale pretrained encoders \autocite{christensen2024echoclipnatmed,vukadinovic2024echoprime,holste2025panecho,zhang2025echovisionfm}.
Self-supervision on \ac{lus} itself would need a \ac{lus} corpus large enough to pretrain on \autocite{vanberlo2023lusssl}, and the one general ultrasound encoder spanning many organs excludes the lung \autocite{jiao2024usfm}.

The closest prior work to the present study is US-JEPA, an image-level \gls{jepa} trained on over five million frames across 22 anatomies including the lung, which asks whether an ultrasound-native latent-predictive model can be trained rather than whether an existing echocardiography model can be repurposed \autocite{radhachandran2026usjepa}.
Its weights are not released, so it cannot enter a comparison here.

\subsection{Cross-domain transfer}
\label{sec:background:transfer}

Transfer into medical imaging has a cautionary history.
\Textcite{raghu2019transfusion} found that ImageNet transfer offered little benefit on two large medical imaging tasks and that far smaller models performed comparably, so an ImageNet initialisation is a reference point rather than a strong prior.
Self-supervised pretraining changes this, matching strong supervised comparators on a fraction of the labelled data \autocite{azizi2023remedis}, yet validation under shifts of anatomy and imaging technique remains inconsistently documented \autocite{vergara2026vfmradiologyreview}, and that is the shift imposed here.

What echocardiography and \ac{lus} share is the physics of image formation rather than the content of the image.
Most soft tissue is structured too finely for an ultrasound scanner to resolve, so its echoes interfere to produce speckle.
The statistics of that speckle are determined by the ultrasound probe geometry and acquisition system, rather than by the organ \autocite{wagner1983specklestatistics}.
An aerated lung departs from this, because air hinders the propagation of ultrasound through the parenchyma, so \ac{lus} is still read mostly from imaging artefacts such as A-lines and B-lines \autocite{demi2023lusguidelines}.
Only where consolidation replaces the air does the lung take on a \textit{tissue-like} echotexture (see \cref{fig:lus}), the appearance closest to that of soft tissue in echocardiogram images.
Ultrasound encoders pretrained across many organs have nonetheless been shown to generalise across organs and tasks \autocite{jiao2024usfm}.

The encoder is held fixed throughout this study.
\Textcite{kumar2022finetuningdistort} show that updating every parameter can distort the pretrained features and leave the model worse than training only a classifier above a fixed encoder, when the target domain lies far from the pretraining domain.
The labelled cohort here is small and the shift in anatomy and acquisition protocol is large, so only the classifier is trained.

Far less attention has been paid to the transformation that stands between a fixed encoder and the classifier.
Because no pretraining objective constrains the scale or spread of the features an encoder produces, a fixed encoder can hand the classifier a representation whose statistics suit it badly.
Whether a shortfall is then a property of the representation or of its scale is what feature conditioning seeks to resolve.

\subsection{Multiple-instance learning for patient-level scoring}
\label{sec:background:mil}

The imaging record of a patient is produced by a clinician sweeping the transducer over a fixed set of anatomical sites on the chest, under a standardised protocol \autocite{suttels2022imagebanks}.
Each site contributes one or more still images to the record of that patient.
A reader interpreting the record looks for the sonographic findings that tuberculosis can produce \autocite{moller2024usfindingseptb}, consolidation among them (see \cref{fig:lus}).
Those findings need \emph{not} be present at every site imaged in a patient who has the disease.
A patient carries a single diagnosis, whereas none of the images in the record of that patient carries a label of its own.
\Ac{mil} matches this structure by presenting the model with an \textit{unordered set} of images, from which it must produce a single \ac{tb} score for the patient \autocite{ilse2018attentionmil}.

Attention-based \ac{mil} learns a weight for every image and pools the image embeddings into a single representation as a weighted mean \autocite{ilse2018attentionmil}.
Because the weights depend on the embeddings rather than on the order of the images, the operation is permutation invariant and accepts sets of different sizes, which a variable number of images per patient requires.
It also carries no assumption that every image in a positive set shows disease, so the model is not required to observe disease-related findings at every anatomical site of a patient confirmed to have \ac{tb}.

\section{Data}
\label{sec:data}

\begin{table}[tb]
	\centering
	\caption{Composition of the lung ultrasound cohort.
	Folds are patient-disjoint, and the sixth is held out as the test set.
	}
	\label{tab:cohort}
	\begin{threeparttable}
		\begin{tabular}{@{}lr@{}}
			\toprule
			Property & Value \\ \midrule
			\multicolumn{2}{@{}l}{\itshape Cohort} \\
			Patients with a label            & 504 \\
			Patients with cleaned images     & 502 \\
			Cleaned images                   & \num{13458} \\
			Images per patient, median       & 28 \\
			Anatomical site codes            & 14 \\
			Cross-validation folds           & 6 \\
			Patients per development fold    & \numrange{79}{81} \\
			Negatives per development fold   & \numrange{41}{46} \\ \midrule
			\multicolumn{2}{@{}l}{\itshape Positive class} \\
			Positive                          & 224 \\
			\quad Bacteriologically confirmed & 192 \\
			\quad Clinically diagnosed        & 32 \\
			Positive fraction                 & \qty{44.6}{\percent} \\
			Held-out test set, positive       & 43 \\
			\quad Bacteriologically confirmed & 38 \\
			\quad Clinically diagnosed        & 5 \\
			Held-out test set, negative       & 58 \\ \bottomrule
		\end{tabular}
\end{threeparttable}
\end{table}

\begin{figure}[!tb]
	\centering
	\captionsetup{subrefformat=parens}
	\begin{subfigure}[t]{0.48\columnwidth}
		\centering
		\includegraphics[width=\linewidth]{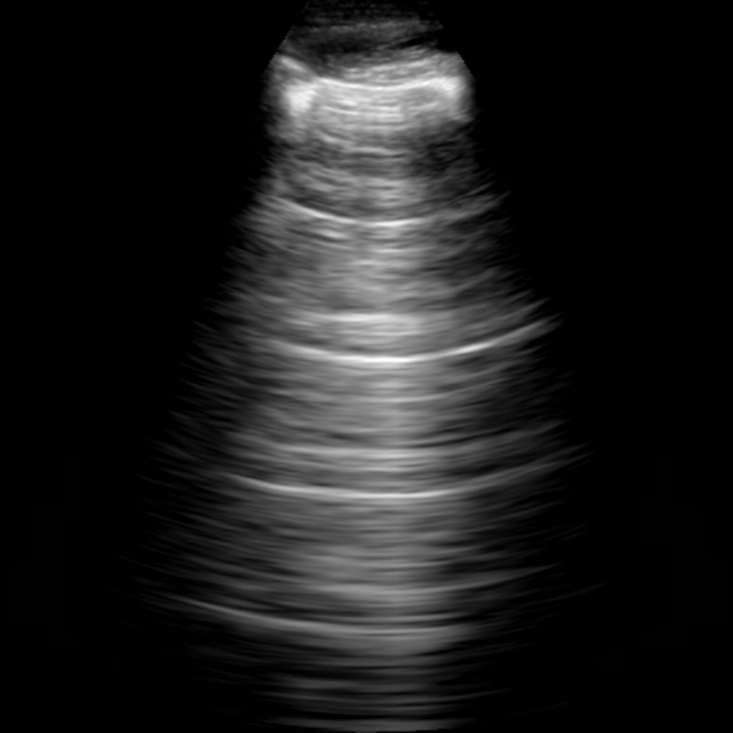}
		\caption{Aerated lung}
		\label{fig:lus:aerated}
	\end{subfigure}
	\hfill
	\begin{subfigure}[t]{0.48\columnwidth}
		\centering
		\includegraphics[width=\linewidth]{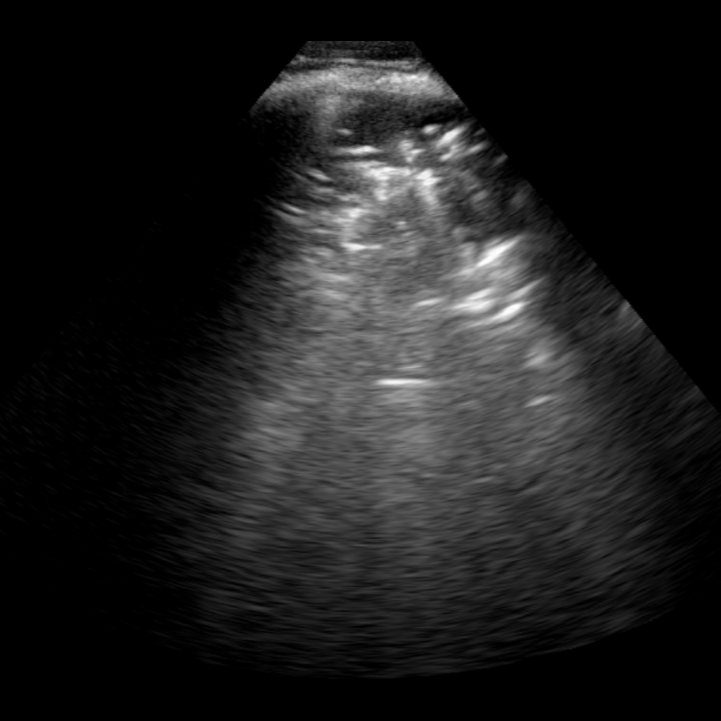}
		\caption{Consolidation}
		\label{fig:lus:consolidation}
	\end{subfigure}
	\caption{Two cleaned images from the target cohort, both acquired at the same
		anterior apical site. In \subref{fig:lus:aerated} a bright pleural line
		lies above a stack of equidistant horizontal reverberation artefacts,
		the appearance of aerated lung. In \subref{fig:lus:consolidation} the
		region beneath the pleural line carries a tissue-like echotexture
		broken by punctate hyperechoic foci, the appearance of consolidation.
		Both patients carry a positive \acs{ptb} label. A patient
		with confirmed disease can therefore present an aerated appearance at
		any individual site, which is what the patient-level aggregation is
		designed to accommodate~(see \cref{sec:setup:classifier}).
	}
	\label{fig:lus}
\end{figure}

The target cohort (see \cref{tab:cohort}) is recruited at a tertiary referral centre in Benin, West Africa, following the same clinical pathway as the baseline system \autocite{ultrai2025escmid}.
Each patient carries a single \ac{ptb} label at the patient level rather than at the level of any image.
A patient counts as positive under either \ac{who} case definition for \ac{tb}, bacteriologically confirmed or clinically diagnosed \autocite{who2013definitions}. Patients may lack the bacteriological result the imaging signal is asked to recover.
Imaging is acquired with a Butterfly iQ+ portable ultrasound probe connected to a smartphone at 14 defined anatomical sites per patient, following the standardised sliding-scan protocol proposed for collaborative \ac{lus} image banks in tuberculosis \autocite{suttels2022imagebanks}.
Every raw frame is \emph{cleaned} by a ruler-anchored crop that removes the device chrome, leaving an 8-bit greyscale image that retains the full sector rather than a tight crop of it (see \cref{fig:lus}).
These cleaned images from the recruited cohort form the dataset used in all experiments that follow.

The echocardiography encoders under test were pretrained on MIMIC-IV-ECHO, a single-centre corpus of transthoracic echocardiography studies \autocite{gow2023mimicivechophysionet}.

\section{Experimental Setup}
\label{sec:setup}

\begin{figure}[!tb]
	\centering
	\includegraphics[width=\columnwidth]{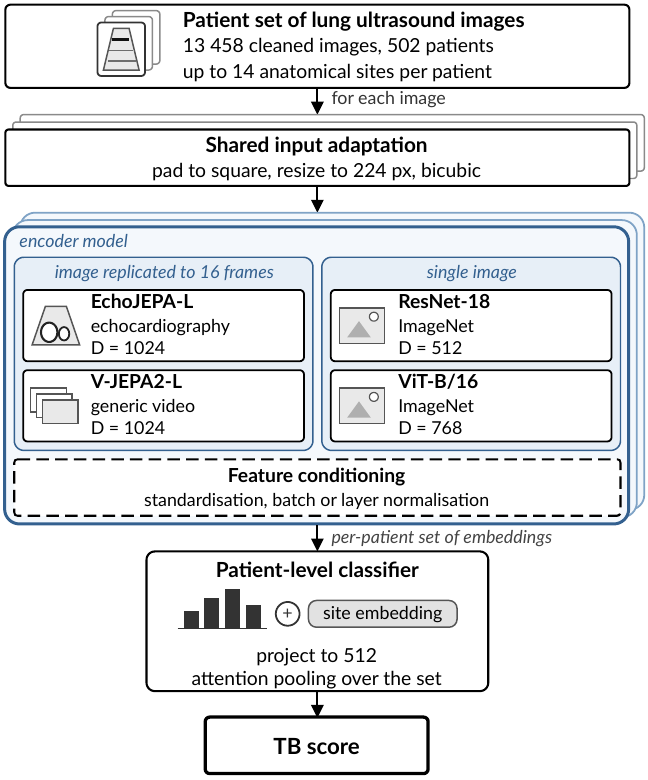}
	\caption{Experimental procedure, in which only the encoder varies.
		Each patient contributes an unordered set of cleaned \acs{lus} images, one per anatomical site, each padded to square, resized to \qtyproduct{224 x 224}{\px} and encoded independently.
		A latent-predictive video encoder instead receives the image replicated to \num{16} identical video frames.
		The features may pass through an optional conditioning stage, and attention pooling with an additive site embedding aggregates the set into a single \acs{tb} score.
		The four encoders drawn here stand for the seventeen compared, each evaluated with its parameters held fixed.
	}
	\label{fig:pipeline}
\end{figure}

\Cref{fig:pipeline} sets out our experimental procedure.
Every stage of it is identical across the seventeen encoders under test, and only the encoder is exchanged.
A measured difference is therefore attributable to encoder architecture and pretraining setup rather than to preprocessing, classifier design or tuning effort.
The encoder parameters are held fixed throughout, so that only the classifier is trained.

\subsection{Input preprocessing and encoders}
\label{sec:setup:encoders}

The cleaned \ac{lus} still images of \cref{sec:data} reach every encoder through one shared preprocessing pipeline.
Each image is padded to a square canvas before resizing, to preserve its aspect ratio.
The padded image is resized to \qtyproduct{224 x 224}{\px} with the bicubic kernel, as done in the baseline system.
Every encoder outside the latent-predictive video group takes that image directly.

The latent-predictive video encoders expect a video rather than a still image, so the image is replicated along the temporal axis to \num{16} identical video frames, as was done for still images during their pretraining.
The token sequence therefore carries no variation along the temporal axis, and the temporal attention operates over identical tokens.
For these video encoders the per-image embedding is the mean over the output tokens.

Each of the seventeen encoders falls into one of four groups: convolutional networks supervised on natural images, transformers supervised on natural images, transformers self-supervised on natural images, and latent-predictive video encoders.

The convolutional encoder group includes \gls{resnet}[-18], -34, -50, -101 and -152~\autocite{he2016resnet}.
All five use the weights obtained by supervised training on ImageNet-1k, with $D=512$ for the first two architectures and $D=2048$ for the remaining three, where $D$ is the dimensionality of the feature vector that each network produces for one input image.
The baseline system uses the \gls{resnet}[-18] encoder, which achieved the best classification performance in prior work on this dataset~\autocite{ultrai2025escmid}.
Every encoder, including this baseline, is optimised over the same hyperparameter range~(see \cref{sec:setup:protocol}).

The supervised transformer encoder group includes \gls{vit}-B/16 with $D=768$, and \gls{vit}-L/16 with $D=1024$.
Both are pretrained on ImageNet-21k and further fine-tuned on ImageNet-1k~\autocite{dosovitskiy2021vit}.
The self-supervised natural-image encoder group includes a \gls{vit}-L/16 \acs{mae} pretrained on ImageNet-1k~\autocite{he2021mae} and a DINOv2 \gls{vit}-L/14 pretrained on LVD-142M~\autocite{oquab2023dinov2}.
For each of these four transformer models we evaluate two variants, taking the encoder representation either from the \glsentrylong{cls} or from the mean of the patch tokens.

The latent-predictive video encoder group includes V-JEPA2-L, the \gls{vit}[-L] configuration of V-JEPA~2, pretrained on generic video~\autocite{assran2025vjepa2}, and three EchoJEPA-L model variants all pretrained on echocardiography at $D=1024$~\autocite{munim2026echojepa}.
The released EchoJEPA-L is a \gls{vit}[-L] checkpoint pretrained on approximately \num{525000} MIMIC-IV-ECHO videos, which distinguishes it from the larger proprietary EchoJEPA-G models of \textcite{munim2026echojepa} trained on roughly 18 million echocardiography videos.
The first uses the released weights, and is named EchoJEPA-L in \cref{tab:backbone}.
The second omits the annealing stage of that pretraining procedure, and is named \textit{pretraining only}.
The third begins from random initialisation rather than from the generic-video initialisation, and is named \textit{without video initialisation}.
All three are used unmodified, as released by \textcite{munim2026echojepa}.

\Gls{vit}-L/16 under supervision is matched to EchoJEPA-L in embedding width and parameter count, but it is not an exact architecture-matched control.
The two also differ in pretraining corpus, in training objective, in how the input is divided into tokens and in how position is encoded, the supervised \gls{vit}[-L/16] taking two-dimensional patches of one image with learned absolute position embeddings where EchoJEPA-L takes tubelets spanning pairs of frames with a \gls{rope} configuration.
V-JEPA2-L is identical to the released EchoJEPA-L in architecture, tubelet size, patch size and \gls{rope} configuration, and differs from it only in the pretraining corpus, which is the contrast this study specified in advance.
No encoder is pretrained on lung ultrasound, or on any ultrasound domain other than echocardiography.

\subsection{Patient-level classifier}
\label{sec:setup:classifier}

Every encoder provides features to the same patient-level classifier.
A \ac{tb} label is recorded per patient, and the set of images belonging to that patient is formed by passing each image independently through the encoder.
The \ac{mil} classifier takes that set and produces one posterior class probability, which serves as the \ac{tb} score of the patient (see \cref{sec:background:mil}).
Each feature vector obtained from the encoder, of width $D = \numrange{512}{2048}$, is linearly projected to a common encoding dimension of \num{512} and summed with a learned embedding associated with each anatomical site.
The linear projection allows encoders of different native width to be compared with an identical \acs{mil} classifier architecture.
As in the baseline system, an attention-pooling \acl{mlp} supplies the weights over the set, and a $512 \rightarrow 1024 \rightarrow \tanh \rightarrow 1$ projection maps the pooled vector to one \ac{tb} score per patient (see \cref{fig:pipeline}).
A padding mask accommodates the variable number of images per patient.
All \num{2.9}~million classifier parameters beyond the input projection are identical across encoders, and no site-dropout augmentation is applied, consistent with the regime of the baseline system.

\subsection{Training and hyperparameter selection protocol}
\label{sec:setup:protocol}

The cohort of \cref{sec:data} is partitioned into six patient-disjoint folds.
The sixth is the held-out test partition, taken unchanged from the baseline system so that the comparison is made on the patients that system was itself held out against~\autocite{ultrai2025escmid}.
The remaining five folds repartition the 401 development patients jointly on the \ac{tb} label and on clinical and radiographic covariates, so that every fold carries a proportional share of each stratum.
For each development fold in turn, the classifier trains on the remaining four and is scored on the development fold left out.
Training runs for \num{20} epochs at an effective batch size of \num{32} patients, using the AdamW optimiser~\autocite{loshchilov2019adamw}, with gradient clipping at unit norm.
The optimiser hyperparameters (learning rate, weight decay and positive-class weight) are searched over \num{45} configurations per encoder within each feature condition, being three learning rates, five weight decays and three positive-class weights.
The configuration and the stopping epoch are chosen by minimising the mean development loss across the five folds, and this is the hyperparameter selection criterion used throughout.
Every figure reported below is a single training run at one seed, except for a repeatability check that repeats four encoders over five seeds and measures how far reinitialisation and shuffle order alone move a result (see \cref{sec:discussion:resolution}).

\subsection{Research questions}
\label{sec:setup:comparisons}

We consider the following two research questions.
In each case, one aspect of the overall system is varied, while all other aspects remain fixed.

\textbf{RQ1)} \emph{Does any encoder improve on the baseline ResNet-18?}
All seventeen encoders are evaluated with their parameters held fixed, so that only the classifier is optimised.
The comparison asks whether any encoder improves on the \gls{resnet}[-18] configuration of the baseline system (see \cref{sec:results:encoders}).

\textbf{RQ2)} \emph{Do fixed encoder features require conditioning before classification?}
The features from each encoder are passed to the classifier in four ways: (1) left unchanged, (2) standardised per dimension to zero mean and unit variance using statistics estimated across the training data, (3) through a batch normalisation layer~\autocite{ioffe2015batchnorm}, and (4) through a layer normalisation layer~\autocite{ba2016layernorm} (see \cref{sec:results:norm}).

\subsection{Metrics, estimand and statistical treatment}
\label{sec:setup:metrics}

The reported quantity throughout is \ac{auc}, and a difference of two areas is reported in signed percentage points.
On the development folds each encoder is summarised by the mean and between-fold standard deviation across the five development-fold models.

On the held-out test set the five development-fold models ensemble the same 101 patients, their scores are averaged per patient, and one test \ac{auc} is computed with a DeLong confidence interval~\autocite{delong1988}.
Because every encoder scores the same patients, each held-out contrast is paired, whether an encoder against the baseline system or the two latent-predictive video encoders against each other.
No comparison ranks one encoder against the rest, so every $p$ value is reported without adjustment for multiplicity.

The development folds support one pre-specified test.
Feature standardisation is applied to every encoder in turn on the same patients, folds and classifier, so the seventeen encoders are paired observations of one intervention, and their differences admit a Wilcoxon signed-rank test~\autocite{wilcoxon1945}.\footnote{A paired $t$ test is not used, the paired differences being strongly right-skewed over two orders of magnitude, a shape for which \textcite{demsar2006} recommends a rank-based test. Every development-fold test is reported two-sided, and the remaining normaliser comparisons are exact sign tests over the same seventeen encoders.}
The encoders share patients, folds, classifier and tuning effort, so what carries the result is the breadth of the family over which the direction holds rather than the exponent.

The power available on the held-out test set is limited.
With 43 positives and 58 negatives, a paired difference in \ac{auc} carries a standard error near \numrange{2}{3} percentage points, so a held-out comparison may fail to reach statistical significance even where the development evidence is strong.

The 2025 screening profile of the \ac{who} sets Type~2 minimal targets of \qty{90}{\percent} sensitivity and \qty{60}{\percent} specificity \autocite{who2025screeningtpp}, for which this work is suited~(see \cref{sec:background:tb}).
We use the \ac{tpp} as a scale against which a measured difference is judged material, not as a certification target.
Specificity at a fixed sensitivity is a step function of the number of negatives, and one patient is the smallest step it can take, so a statistically detectable gain in \ac{auc} can be clinically unresolvable on a cohort of this size (see \cref{sec:discussion:resolution}).

\section{Results}
\label{sec:results}

We first retrain the baseline system on the data set up of this study, to enable a direct numerical comparison~(see \cref{sec:data}).
Results then follow the two research questions in turn, each reported as the mean improvement in \ac{auc}, its bearing on the Type~2 targets, and which contrasts reach statistical significance.

\subsection{The baseline system}
\label{sec:results:baseline}

The baseline system described in \autocite{ultrai2025escmid} is retrained under the protocol described in \cref{sec:setup:protocol}.
This baseline achieves a mean development \ac{auc} of \num{0.8717}, with a between-fold standard deviation of \num{0.0389}.
On the held-out test set, the same system achieves an \ac{auc} of \num{0.8881} $[0.8209, 0.9554]$.\footnote{These figures differ slightly to those published in \autocite{ultrai2025escmid}, where the positive class consisted only of bacteriologically confirmed TB cases.}

\begin{table*}[t]
	\centering
	\caption{Seventeen encoders on the five development folds, each evaluated with and without feature standardisation.
	}
	\label{tab:backbone}
	\begin{threeparttable}
		\begin{tabular}{@{}ll
			S[table-format=1.4(3)]
			S[table-format=1.4(3)]
			S[table-format=+1.2, round-precision=2, round-pad=true,
			  drop-zero-decimal=false]@{}}
			\toprule
			& & \multicolumn{2}{c}{Dev.\ \acs{auc} (sd)} & \\
			\cmidrule(lr){3-4}
			Encoder & Pretraining corpus & {Raw features} & {Standardised features} & {$\Delta$} \\ \midrule
			Baseline system~\autocite{ultrai2025escmid} & ImageNet-1k & {\num{0.8717} $\pm$ \num{0.0389}} & {\textendash} & {\textendash} \\ \midrule
			\multicolumn{5}{@{}l}{\itshape Convolutional, supervised on natural images} \\
			\acs{resnet}-18\tnote{a}                        & ImageNet-1k        & 0.8694(290)           & 0.8817(198)           & +1.23 \\
			\acs{resnet}-34                                 & ImageNet-1k        & 0.8567(453)           & 0.8680(303)           & +1.13 \\
			\acs{resnet}-50                                 & ImageNet-1k        & 0.8704(409)           & 0.8715(447)           & +0.11 \\
			\acs{resnet}-101\tnote{$\dag$}                  & ImageNet-1k        & \bfseries 0.8759(316) & 0.8835(163)           & +0.76 \\
			\acs{resnet}-152                                & ImageNet-1k        & 0.8741(366)           & \bfseries 0.8871(322) & \bfseries +1.30 \\ \midrule
			\multicolumn{5}{@{}l}{\itshape Transformer, supervised on natural images} \\
			\acs{vit}-B/16                                  & ImageNet-21k to 1k & 0.8545(484)           & 0.8709(497)           & \bfseries +1.64 \\
			\acs{vit}-B/16, mean patch                      & ImageNet-21k to 1k & 0.8719(311)           & 0.8805(313)           & +0.86 \\
			\acs{vit}-L/16\tnote{$\dag$}                    & ImageNet-21k to 1k & 0.8674(271)           & \bfseries 0.8831(259) & +1.57 \\
			\acs{vit}-L/16, mean patch                      & ImageNet-21k to 1k & \bfseries 0.8746(396) & 0.8803(226)           & +0.57 \\ \midrule
			\multicolumn{5}{@{}l}{\itshape Transformer, self-supervised on natural images} \\
			\acs{vit}-L/16 \acs{mae}\tnote{$\dag$}          & ImageNet-1k        & 0.8509(313)           & 0.8881(317)           & \bfseries +3.72 \\
			\acs{vit}-L/16 \acs{mae}, mean patch            & ImageNet-1k        & 0.8610(315)           & \bfseries 0.8922(253) & +3.12 \\
			DINOv2 \acs{vit}-L/14                           & LVD-142M           & 0.8609(399)           & 0.8724(347)           & +1.15 \\
			DINOv2 \acs{vit}-L/14, mean patch               & LVD-142M           & \bfseries 0.8663(278) & 0.8666(485)           & +0.03 \\ \midrule
			\multicolumn{5}{@{}l}{\itshape Latent-predictive video} \\
			V-JEPA2-L\tnote{$\dag$}                         & generic video      & 0.8645(340)           & 0.8796(254)           & \bfseries +1.51 \\
			EchoJEPA-L\tnote{$\dag$}                        & MIMIC-IV-ECHO      & \bfseries 0.8724(264) & \bfseries 0.8803(108) & +0.79 \\
			EchoJEPA-L, pretraining only                    & MIMIC-IV-ECHO      & 0.8694(385)           & 0.8763(218)           & +0.69 \\
			EchoJEPA-L, without video initialisation        & MIMIC-IV-ECHO      & 0.8707(454)           & 0.8774(336)           & +0.67 \\ \bottomrule
		\end{tabular}
		\begin{tablenotes}[flushleft]
			\footnotesize
			\item[] Every entry is the mean development \acs{auc} across folds 1 to
			5, followed by the between-fold standard deviation, both in units of
			\acs{auc}, each taken at the configuration the hyperparameter selection
			criterion chose out of the 45 searched within that condition.
			\item[] $\Delta$ is standardised minus raw, paired within the encoder on
			the same patients, folds and classifier, in signed percentage points.
			The baseline system updates all of its parameters and has no feature
			condition to ablate, so it holds one entry and carries no
			$\Delta$~(see \cref{sec:results:baseline}).
			\item[] \textbf{Bold} marks the leading value of each group under each
			feature condition, and the leading $\Delta$ of each group.
			\item[$\dag$] Carried to the held-out test set~(see \cref{tab:main}).
			\item[a] The encoder the baseline system uses, with its parameters
			held fixed where the baseline row updates them.
		\end{tablenotes}
	\end{threeparttable}
\end{table*}

\begin{table*}[t]
	\centering
	\caption{The held-out test set of 101 patients, 43 positive and 58 negative.
		Every figure is a mean-probability ensemble over the five development-fold models, whose scores are averaged per patient before one \acs{auc} is computed on those 101 patients.
	}
	\label{tab:main}
	\begin{threeparttable}
		\setlength{\tabcolsep}{3pt}
		\footnotesize
\begin{tabular}{@{}l
			S[table-format=1.4]
			I{table-format=1.4}
			S[table-format=+1.2, round-precision=2, round-pad=true,
			  drop-zero-decimal=false]
			I{table-format=+1.2, round-precision=2, round-pad=true,
			  drop-zero-decimal=false}
			S[table-format=1.3, round-precision=3, round-pad=true,
			  drop-zero-decimal=false]
			S[table-format=2.1, round-precision=1, round-pad=true,
			  drop-zero-decimal=false]
			I{table-format=2.1, round-precision=1, round-pad=true,
			  drop-zero-decimal=false}
			S[table-format=+2.2, round-precision=2, round-pad=true,
			  drop-zero-decimal=false]
J{table-format=+2.1, round-precision=1, round-pad=true,
			  drop-zero-decimal=false}@{}}
			\toprule
System or contrast
			& \multicolumn{3}{c}{Test \acs{auc}}
			& \multicolumn{3}{c}{$\Delta$ test \acs{auc}}
			& {$p$}
			& \multicolumn{3}{c}{Specificity (\unit{\percent})}
			& \multicolumn{3}{c}{$\Delta$ spec.} \\
\midrule
Baseline system~\autocite{ultrai2025escmid} & 0.8881 & 0.8209 & 0.9554 & \multicolumn{3}{c}{\textendash} & {\textendash} & 60.3 & 23.6 & 87.1 & \multicolumn{3}{c}{\textendash} \\ \midrule
			\multicolumn{14}{@{}l}{\itshape Encoders against the baseline system, each paired on identical patients} \\
			\acs{vit}-L/16, supervised     & 0.9138 & 0.8555 & 0.9721 & \bfseries +2.57 & +0.12 & +5.01 & \bfseries 0.040 & 74.1 & 25.9 & 91.8 & +13.79 & -6.0  & +27.6 \\
			\acs{vit}-L/16, \acs{mae}      & 0.9118 & 0.8512 & 0.9724 & +2.37 & -0.89 & +5.62 & 0.154 & 79.3 & 26.8 & 92.3 & +18.97 & -8.6  & +33.3 \\
			V-JEPA2-L                      & 0.9058 & 0.8435 & 0.9680 & +1.76 & -0.88 & +4.41 & 0.191 & 69.0 & 28.8 & 91.4 & +8.62  & -17.0 & +24.1 \\
			EchoJEPA-L                     & 0.9042 & 0.8469 & 0.9615 & +1.60 & -2.09 & +5.29 & 0.394 & 70.7 & 39.6 & 88.6 & +10.34 & -8.2  & +27.1 \\
			\acs{resnet}-101               & 0.8905 & 0.8244 & 0.9567 & +0.24 & -2.56 & +3.05 & 0.866 & 56.9 & 31.7 & 86.5 & -3.45  & -20.8 & +21.4 \\
			\acs{resnet}-101, raw features & 0.8777 & 0.8076 & 0.9478 & -1.04 & -4.31 & +2.22 & 0.532 & 58.6 & 36.7 & 87.5 & -1.72  & -21.2 & +25.0 \\ \midrule
			\multicolumn{14}{@{}l}{\itshape Feature conditioning, within \acs{resnet}-101, specified in advance} \\
			Batch-normalised against raw   & \multicolumn{3}{c}{\textendash} & +1.28 & -0.99 & +3.55 & 0.268 & \multicolumn{3}{c}{\textendash} & -1.72 & -13.5 & +23.0 \\ \midrule
			\multicolumn{14}{@{}l}{\itshape The encoder contrast specified in advance, the two compared only with each other} \\
			EchoJEPA-L against V-JEPA2-L   & \multicolumn{3}{c}{\textendash} & -0.16 & -3.54 & +3.21 & 0.926 & \multicolumn{3}{c}{\textendash} & +1.72 & -11.1 & +26.7 \\ \bottomrule
		\end{tabular}
		\begin{tablenotes}[flushleft]
			\footnotesize
			\item[] Bracketed figures are \qty{95}{\percent} confidence intervals,
			and every specificity is measured at \qty{90}{\percent} sensitivity.
			\item[] $\Delta$ \acs{auc} is in signed percentage points from a
			paired \textcite{delong1988} test on identical patients, and
			$\Delta$ spec.\ is in signed percentage points of specificity at
			\qty{90}{\percent} sensitivity from a paired patient bootstrap over
			\num{10000} resamples with the threshold re-estimated in every
			resample. Every comparison is paired between two systems on identical
			patients, so no adjustment for multiplicity applies and every $p$ is
			reported as measured~(see \cref{sec:setup:metrics}).
			\item[] \textbf{Bold} marks each two-sided $p < \num{0.05}$.
		\end{tablenotes}
	\end{threeparttable}
\end{table*}

\subsection{Encoder choice}
\label{sec:results:encoders}

\begin{figure}[!tb]
	\centering
	\includegraphics[width=\columnwidth]{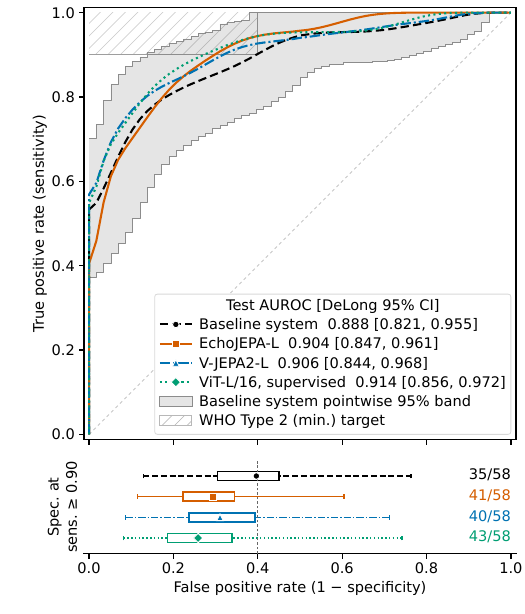}
	\caption{The held-out test set, on the 101 patients of
		\cref{tab:cohort}, for the baseline system, for EchoJEPA-L and
		V-JEPA2-L, and for the encoder that leads there.
		\emph{Upper panel:} each curve is the pointwise mean over
		\num{1000000} bootstrap resamples at the \num{59} achievable
		false-positive rates, and the legend gives the empirical \acs{auc} with
		a DeLong \qty{95}{\percent} interval \autocite{delong1988}. The grey band is
		the pointwise \qty{95}{\percent} bootstrap interval of the baseline
		system alone.
        The
		hatched region marks operating points meeting both Type~2 minimal
		targets of the 2025 screening profile of the \ac{who}, \qty{90}{\percent}\
		sensitivity and \qty{60}{\percent} specificity. \emph{Lower panel:}
		the operating points instead, as specificity at a sensitivity of at
		least \qty{90}{\percent} on the same axis, where the box is the
		interquartile range of the bootstrap, the whisker its
		\qty{95}{\percent} interval with the threshold re-estimated inside every
		resample, the marker the observed value rather than the bootstrap
		median, and the right-hand figures the true negatives of 58. The
		dashed rule is the Type~2 specificity.
	}
	\label{fig:results}
\end{figure}

\Cref{tab:backbone} reports the development results for the seventeen encoders under both raw and standardised feature conditions.
\Cref{tab:main} and \cref{fig:results} give the held-out test results for the encoder models carried forward.

On the development folds the best-to-worst spread across the seventeen encoders is \pp{2.50} percentage points on raw features and \pp{2.56} on standardised features, and no encoder stands more than \pp{+2.05} points above the baseline.
The \acs{resnet}-101 leads on raw features at a mean \ac{auc} of \num{0.8759}, and the \gls{vit}-L/16 \acs{mae} on its mean-patch representation leads on standardised features.
On the held-out test set every encoder selected on the development folds stands above the baseline in \ac{auc}, the supervised \gls{vit}-L/16 by the largest margin at \pp{+2.57} percentage points and the \acs{resnet}-101 by the smallest at \pp{+0.24}.

Contrasting EchoJEPA-L against V-JEPA2-L measures a \pp{-0.16} percentage point reduction on the held-out test set.
This reverses the direction observed on the development folds, where echocardiography pretraining showed an improvement in classification transferability.

Against the Type~2 targets the separation is wider at the operating point.
Where the baseline reaches \qty{60.3}{\percent} specificity at \qty{90}{\percent} sensitivity, four of the five selected encoders reach between \qty{69.0}{\percent} and \qty{79.3}{\percent}, a gain of up to \pp{+18.97} percentage points, while the \acs{resnet}-101 alone falls below the baseline at \qty{56.9}{\percent}.
The hatched region of \cref{fig:results} marks the operating points that meet both Type~2 minima.

Among these held-out contrasts only the supervised \gls{vit}-L/16 against the baseline reaches statistical significance, at $p = \num{0.040}$.
The \acs{resnet}-101 margin returns $p = \num{0.866}$, every paired difference in specificity spans zero, and the EchoJEPA-L against V-JEPA2-L contrast returns $p = \num{0.926}$ with an interval of $[-3.54, +3.21]$ percentage points.

\begin{table}[tb]
	\centering
		\caption{Which normaliser, paired within each encoder.
		Each row names two conditions and reports the first against the second, so a positive $\Delta$ favours the first.
		}
		\label{tab:norm}
	\begin{threeparttable}
\setlength{\tabcolsep}{2pt}
		\footnotesize
\begin{tabular}{@{}lc
			S[table-format=+1.2, round-precision=2, round-pad=true,
			  drop-zero-decimal=false]
			I{table-format=+1.2, round-precision=2, round-pad=true,
			  drop-zero-decimal=false}
			c@{}}
			\toprule
Comparison & {Ahead} & \multicolumn{3}{c}{$\Delta$ dev.\ \acs{auc}}
			& {$p$} \\ \midrule
			\multicolumn{6}{@{}l}{\itshape Does feature conditioning improve downstream classification?} \\
			Standard. vs raw            & 17/17 & +1.23 & +0.83 & +1.70 & \num{1.5e-5} \\
			Batch norm.\ vs raw         & 15/17 & +1.12 & +0.67 & +1.65 & \num{0.0023} \\
			Layer norm.\ vs raw         & 4/17  & -0.35 & -0.66 & -0.02 & \num{0.049} \\ \midrule
			\multicolumn{6}{@{}l}{\itshape Should features be normalised across the dataset, or merely learned?} \\
			Standard. vs layer norm.    & 17/17 & +1.57 & +1.16 & +2.00 & \num{1.5e-5} \\
			Batch norm.\ vs layer norm. & 17/17 & +1.46 & +1.04 & +1.92 & \num{1.5e-5} \\ \midrule
			\multicolumn{6}{@{}l}{\itshape Can feature standardisation be substituted with batch normalisation?} \\
			Batch norm.\ vs standard.   & 8/17  & -0.11 & -0.53 & +0.31 & \num[round-precision=1, round-pad=true, drop-zero-decimal=false]{1.000} \\ \bottomrule
		\end{tabular}
		\begin{tablenotes}[flushleft]
			\footnotesize
			\item[] Every entry is a mean development \acs{auc} over all 17
			encoders. \emph{Ahead} counts the encoders on which the first named
			normaliser leads.
			\item[] $\Delta$ is in signed percentage points, given as the mean
			followed by a bracketed \qty{95}{\percent} interval, which is a
			percentile bootstrap over \num{10000} resamples of the 17 encoders.
			\item[] Every $p$ is two-sided, the first row from a Wilcoxon
			signed-rank test and the rest from exact sign
			tests~(see \cref{sec:setup:metrics}).
		\end{tablenotes}
	\end{threeparttable}
\end{table}

\subsection{Feature conditioning}
\label{sec:results:norm}

Standardising the features between the encoder and the classifier improves all seventeen encoders, by a mean of \pp{+1.23} percentage points with a bootstrap interval of $[+0.83, +1.70]$, which is \qty{48}{\percent} of the best-to-worst spread across encoders (see \cref{tab:norm}).
On the held-out test set the one conditioning contrast carried, batch-normalised against raw features within \acs{resnet}-101, gains \pp{+1.28} percentage points of \ac{auc}.

The development gain is statistically significant, at a Wilcoxon $p = \num{1.5e-5}$.
The two normalisers that estimate statistics across the dataset --- standardisation and batch normalisation --- are indistinguishable, differing by \pp{-0.11} percentage points at $p = \num{1.000}$, though failing to detect a difference does not establish that the two are equivalent.
By contrast, layer normalisation, which estimates statistics for each image, falls below raw features at \pp{-0.35} percentage points and $p = \num{0.049}$.

\section{Discussion}
\label{sec:discussion}

\Cref{fig:results} grounds this discussion, placing the baseline, the two latent-predictive video encoders and the leading encoder on the held-out patients against the Type~2 targets.
The question this study set out to answer is not which encoder scores highest, but what constitutes a real improvement on a cohort of this size.
The resolution of the measurement answers it, and the two research questions are read against that resolution below.

The mean development \ac{auc} moves by \pp{7.30} percentage points according to which patients fall in a fold, so twice the standard error of that fold mean, \pp{2.71} percentage points, is the smallest difference between two encoders this study can resolve.
No adjacent pair of encoders in \cref{tab:backbone} clears that figure, and the best-to-worst spread of the whole family sits inside a single unit of that resolution under either feature condition.

\subsection{Encoder choice (RQ1)}
\label{sec:discussion:rq1}

Held fixed and given conditioned features, every encoder selected on the development folds improved on the baseline system on the held-out patients, the two transformers by \pp{+2.57} and \pp{+2.37} percentage points of \ac{auc} and by \pp{+13.79} and \pp{+18.97} points of specificity at \qty{90}{\percent} sensitivity.
A gain of that size at the operating point would matter clinically, and a further cohort would be recruited to confirm it.
Replacing the incumbent \gls{resnet}[-18] encoder with any of them is supported by every selected encoder clearing the baseline, and in a consistent direction.
It is not evidence that one replacement is better than another, because the whole band sits inside what these 101 patients can measure (see \cref{sec:discussion:resolution}).
Replacing the incumbent encoder is therefore supported, whereas choosing among the replacements is a matter of recruitment rather than of method.

In a \ac{tb} screening, due to the low prevalence of the disease, most people tested do not have \ac{tb}, so it is specificity, not sensitivity, that governs the number of false-positive results.
Consider a setting in which 1000 people are screened for \ac{tb} and the prevalence is between \qty{1}{\percent} and \qty{10}{\percent}.
If specificity were raised from \qty{60.3}{\percent} to \qty{79.3}{\percent} at a fixed \qty{90}{\percent} sensitivity, between 171 and 188 fewer people would be sent for a second screening test.
Under a Type~2 pathway a false positive leads to a further screening test rather than to treatment, so its cost falls on confirmatory testing, laboratory capacity and time lost by the patient (see \cref{sec:background:tb}).
These are point estimates from a single tertiary-referral cohort, and the specificity gain behind them is not resolved by these data, so they show the possible scale of the benefit rather than a reduction to expect in practice.

The contrast the study was built on, echocardiography against generic-video pretraining, returns no statistical advantage for echocardiography.
EchoJEPA-L against V-JEPA2-L lies within resolution of the development folds and changes sign on the held-out test set, a direction \textcite{taratynova2025cardiobench} also report inside the cardiac domain.
The interval is too wide to establish equivalence, so the finding is not that the benefit is small but that any benefit lies below what this cohort can measure.

That bound is moreover for a video encoder presented with replicated still images rather than with a recording, which the blind-sweep protocol does not produce (see \cref{sec:data,sec:setup:encoders}), so the study cannot separate a representation that carries nothing useful here from one that the replicated video frames degraded.
The physics of image formation offers one reason the first could hold, since \ac{lus} is read largely from artefacts whereas echocardiography is read from images of tissue (see \cref{sec:background:transfer}).
The follow-up that would separate them is a cohort acquired as recordings.

\subsection{Feature conditioning (RQ2)}
\label{sec:discussion:rq2}

What stands between a fixed encoder and the classifier matters more than which encoder it is.
The comparison establishes that \emph{where} the feature statistics are estimated is what matters most, rather than whether a learned transform accompanies them.
The two normalisers that estimate across the dataset perform alike, whether the statistics are fitted on the training split beforehand or accumulated from mini-batches as training proceeds.
Layer normalisation showed no such improvement over raw features, as it remains confined to one image at a time.
A trainable affine transform is therefore not the active ingredient, whereas the application of dataset-wide scale information is.
Batch normalisation is the more practical of the two effective normalisers, since it estimates its statistics during training and needs no separate pass over the dataset beforehand.

\subsection{Measurement resolution and limitations}
\label{sec:discussion:resolution}

The measurement resolution is a floor rather than an estimate, since the folds are not exchangeable and one is the hardest and another the easiest for most encoders.
Reseeding alone moves an encoder by \numrange{0.28}{0.68} percentage points, and one encoder selects five different configurations across five seeds while its mean development \ac{auc} moves by only \pp{0.70} percentage points.
For that encoder the configuration is better read as a lottery over a flat region than as a stable property.
Only the standard error of the fold mean bears on whether two encoders differ, most of the fold-sampling noise cancelling in a difference taken on the same encoder over the same folds.

The clinical unit is coarser still, and this is where the size of the cohort matters most.
Specificity at a fixed sensitivity is a step function of the number of negatives, and one patient is the smallest step it can take, \numrange{2.2}{2.4} percentage points on a development fold and \pp{1.72} on the held-out test set.
An effect below that line is one no operating point on this cohort can register.
At \qty{90.7}{\percent} sensitivity the baseline system reaches \qty{60.3}{\percent} specificity, with a patient bootstrap of $[23.6, 87.1]$, which places it on the Type~2 minimum.
However, every interval in the lower panel of \cref{fig:results} crosses that minimum.
This cohort therefore does not resolve which side of the Type~2 specificity any of these systems sits on.

The cohort also bounds how far these results generalise.
Every patient is recruited at one tertiary referral centre and imaged with one portable probe under one standardised protocol, and the held-out partition comes from that same site, so the held-out evaluation is an internal validation rather than an external one.
A referral population at a positive fraction of \qty{44.6}{\percent} is also likely to differ in disease spectrum from people presenting at primary care, so sensitivity and specificity may both shift in screening use.

\section{Summary and Conclusion}
\label{sec:conclusion}

We compared seventeen pretrained encoders for \ac{lus} tuberculosis screening under a protocol in which only the encoder varied, each evaluated with its parameters held fixed.

Feature conditioning improved all seventeen encoders and is the one intervention the development folds establish.
On the held-out test set every encoder selected on those folds placed above the baseline system, the largest margins being \pp{+2.57} percentage points of \ac{auc} and \pp{+18.97} percentage points of specificity at \qty{90}{\percent} sensitivity.
Choosing among those encoders is what this cohort cannot settle, every one of them falling inside its measurement resolution.

Reusing a high-resource ultrasound encoder in a low-resource ultrasound domain remains worth attempting, and it was not what limited this study.
Shared imaging physics did not on its own make echocardiography a productive pretraining corpus, and no advantage for it was detected under static-frame evaluation.
Whether such a system meets the WHO Type~2 minimum is a question of recruitment rather than of encoder choice.

\section*{Ethics and Data Availability}
\label{sec:ethics}

The study that recruited the target cohort was approved by the committee for biomedical research ethics of the University of Parakou on 18 May 2021, under reference 0407/CLERB-UP/P/SP/R/SA, and is registered as NCT05423847 \autocite{ultrai2025escmid}.
Every participant gave written informed consent, and a trusted representative assisted any participant who was unable to read or write.

The de-identified images, reproduced in \cref{fig:lus}, and their patient-level labels are deposited under a persistent identifier \autocite{lusdataset2025}.
Access is granted for research use under a data use agreement.

The echocardiography and generic-video encoders are those released by their respective authors and are used unmodified \autocite{munim2026echojepa,assran2025vjepa2}.
MIMIC-IV-ECHO, the corpus behind the echocardiography encoders, is available under credentialed access, governed by the PhysioNet Credentialed Health Data License 1.5.0 and the corresponding data use agreement \autocite{gow2023mimicivechophysionet}.
No echocardiography frame is processed or reproduced in this work.

\section*{Acknowledgements}

\begin{hidden}[\textit{Acknowledgements omitted for purpose of blind peer-review}]
This project has received funding from the European Union’s Horizon Europe research and innovation Programme Global Health EDCTP3 Joint Undertaking under grant agreement No. 101190662 -- CAD LUS4TB, and has received funding from the Swiss State Secretariat for Education, Research and Innovation (SERI) under contract number 589188.
The TrUST study, within which the target cohort was recruited, is funded by the Swiss Lung Association.

In addition, we gratefully acknowledge financial support from Telkom (South Africa) for the research presented in this paper.
We also thank the Stellenbosch Rhasatsha High Performance Computing facility and team for access to their facilities and for their technical support, which were invaluable to this project.

The Butterfly iQ+ ultrasound equipment used in this project was provided by BFLY Operations, Inc.
BFLY Operations, Inc. had no involvement in the funding, data collection, design, analysis and outcome related to this project.
\end{hidden}

\section*{ORCID}
\begin{hidden}[\textit{ORCID identifiers omitted for purpose of blind peer-review}]
\small
\newcommand{\orcidline}[2]{\textit{#1} & \href{https://orcid.org/#2}{\includegraphics[width=1em]{orcid.pdf} orcid.org/#2}\\}
\noindent\begin{tabular*}{\linewidth}{@{}l@{\extracolsep{\fill}}r@{}}
\orcidline{Christiaan M. Geldenhuys}{0000-0003-0691-0235}
\orcidline{Joshua M. Jansen van Vüren}{0000-0002-3406-4788}
\orcidline{Véronique Suttels}{0000-0002-5261-0439}
\orcidline{Trevor Brokowski}{0000-0002-7547-7749}
\orcidline{Ablo P. Wachinou}{0000-0002-1024-9665}
\orcidline{Mary-Anne Hartley}{0000-0002-8826-3870}
\orcidline{Rensu P. Theart}{0000-0003-0508-1690}
\orcidline{Grant Theron}{0000-0002-9216-2415}
\orcidline{Thomas R. Niesler}{0000-0002-7341-1017}
\end{tabular*}
\end{hidden}

\printbibliography

\end{document}